\documentclass[12pt]{article}%
\usepackage{graphicx}
\usepackage[english]{babel}
 
\begin{document}

\begin{center}{\large\bf Cosmological Acceleration as a Consequence of Quantum de Sitter Symmetry} \end{center}

\vskip 1em \begin{center} {\large Felix M. Lev} \end{center}
\vskip 1em \begin{center} {\it Artwork Conversion Software Inc.,
509 N. Sepulveda Blvd, Manhattan Beach, CA 90266, USA
(Email:  felixlev314@gmail.com)} \end{center}

\begin{abstract}
Physicists usually understand that physics cannot (and should not) derive that $c\approx 3\cdot 10^8m/s$ 
and $\hbar \approx 1.054\cdot 10^{-34}kg\cdot m^2/s$. At the same time they usually believe that physics should derive the value of the cosmological constant $\Lambda$ and that the solution of the dark energy problem depends on this value. However, background space in General Relativity (GR) is only a classical notion while on quantum level symmetry is defined by a Lie algebra of basic operators. We prove that
the theory based on Poincare Lie algebra is a special degenerate case of the theories based on de Sitter 
(dS) or anti-de Sitter (AdS) Lie algebras in the formal limit $R\to\infty$ where R is the parameter of contraction from the latter algebras to the former one, and $R$ has nothing to
do with the radius of background space. As a consequence,  $R$ is necessarily finite, is fundamental to the same extent as
$c$ and $\hbar$, and a question why $R$ is as is does not arise. Following our previous publications, we consider a system of two free bodies in dS quantum mechanics and show that in semiclassical approximation the cosmological dS acceleration is necessarily nonzero and is the same as in 
GR if the radius of dS space equals $R$ and $\Lambda=3/R^2$. This result follows from basic principles of quantum theory. It has nothing to do with existence or nonexistence of dark energy
and therefore for explaining cosmological acceleration dark energy is not needed. The result is obtained without using the notion of dS background space 
(in particular, its metric and connection) but simply as a consequence of quantum mechanics based on
the dS Lie algebra. Therefore, $\Lambda$ has a physical meaning only on classical level and the cosmological constant problem and the dark energy problem do not arise. In the case of dS and AdS symmetries all physical quantities are dimensionless and no system of units is needed. In particular, the quantities $(c,\hbar,s)$, which are the basic quantities in the
modern system of units, are not so fundamental as in relativistic quantum theory. "Continuous time" is a part of classical notion of space-time continuum and makes no sense beyond this notion. In particular, description of the inflationary stage of the Universe by times $(10^{-36}s,10^{-32}s)$ has no physical meaning.
\end{abstract}

\begin{flushleft} Keywords: dark energy, quantum theory, de Sitter symmetry\end{flushleft}

\section{Brief overview of the cosmological constant problem and dark energy}
\label{intro}

The history of General Relativity (GR) is described in a vast literature. The Lagrangian of GR is linear in Riemannian curvature $R_c$, but from the point of view of symmetry requirements there exist infinitely many
Lagrangians satisfying such requirements. For example, $f(R_c)$ theories of gravity are widely discussed, where there can be many possibilities for choosing the function $f$.
 Then the effective gravitational
constant $G_{eff}$ can considerably differ from standard gravitational constant $G$. It is also argued that GR is a low energy approximation of more general theories involving higher order derivatives. The nature of gravity on quantum level is a problem, and standard canonical quantum gravity is not renormalizable. For those reasons
the quantity $G$ can be treated only as a phenomenological parameter but not fundamental
one.

Let us restrict ourselves with the consideration of standard GR. Here the Einstein equations depend on two arbitrary parameters $G$ and $\Lambda$ where $\Lambda$ is the cosmological constant (CC). In the formal limit when matter disappears, space-time becomes Minkowski space when $\Lambda=0$, de Sitter (dS) space when $\Lambda>0$, and anti-de Sitter (AdS) space when $\Lambda<0$.

Well known historical facts are
that first Einstein included $\Lambda$ because he believed that the Universe should be stationary, and this is possible only if $\Lambda \neq 0$. However, according to Gamow,  after Friedman's results and Hubble's discovery of the Universe
expansion, Einstein changed his mind and said that inclusion of $\Lambda$ was the greatest blunder of his life. 

The usual philosophy of GR is that curvature is created by matter and therefore
$\Lambda$ should be equal to zero. This philosophy has been advocated even
in standard textbooks written before 1998. For example, the authors of Ref. \cite{LL}
say that "...there are no convincing reasons, observational and theoretical, for introducing
a nonzero value of $\Lambda$" and that "... introducing to the density of the Lagrange function a constant term which does not depend on the field state would mean attributing to space-time a principally ineradicable curvature which is related neither to matter nor to gravitational waves".

However, the data of Ref. \cite{Perlmutter} on supernovae have shown that $\Lambda > 0$ with the accuracy better than $5\%$, and further investigations have improved the accuracy to $1\%$. For reconciling this fact with the philosophy of GR, the terms with 
$\Lambda$ in the left-hand-sides of the Einstein equations have been moved to the 
right-hand-sides and interpreted not as the curvature of empty space-time but as a
contribution of unknown matter called dark energy. Then, as follows from the experimental value of $\Lambda$, dark energy contains approximately $70\%$ of the
energy of the Universe. At present a possible nature of dark energy is discussed in a
vast literature and several experiments have been proposed.

Let us note the following. In the formalism of GR the coordinates
and curvature are needed for the description of real bodies. One of fundamental principles of physics is that definition of a physical quantity is the description on how this quantity should be measured. In the Copenhagen formulation of quantum theory measurement is an interaction with a classical object. Therefore since in empty space-time nothing can be measured, the coordinates and
curvature of empty space-time have no physical meaning. This poses a problem whether
the formal limit of GR when matter disappears but space-time remains is physical. Some authors (see e.g. Ref. \cite{Hikin}) propose approaches such that if matter disappears then space-time disappears too. 

The CC problem is as follows. In standard quantum field theory one starts from the choice of the space-time background. By analogy with the philosophy of GR,
it is believed that the choice of the Minkowski background is more physical than the choice of the dS or AdS one. Here the quantity $G$ is treated as fundamental and 
the value of $\Lambda$ should be extracted from the vacuum expectation value of
the energy-momentum tensor. The theory contains strong
divergencies and a reasonable cutoff gives for $\Lambda$ a value exceeding
the experimental one by 120 orders of magnitude. This result is expected because in
units $c=\hbar=1$ the dimension of $G$ is $m^2$, the dimension of $\Lambda$ is
$m^{-2}$ and therefore one might think than $\Lambda$ is of the order of $1/ G$
what exceeds the experimental value by 120 orders of magnitude. 

Several authors argue that the CC problem does not exists. For example, the
authors of Ref. \cite{Bianchi} titled "Why all These Prejudices Against a Constant?"
note that since the solution of the Einstein 
equations depends on two {\it arbitrary phenomenological}  constants $G$ and 
$\Lambda$ it is not clear why we should choose only a special case $\Lambda=0$.  
If $\Lambda$ is as small as given in Ref. \cite{Perlmutter} then
it has no effect on the data in Solar System and the contribution of $\Lambda$ is
important only at cosmological distances. Also theorists supporting Loop Quantum Gravity
say that the preferable choice of Minkowski background contradicts the background
independence principle. Nevertheless, the majority of physicists working in this
field believe that the CC problem does exist and the solution should be sought in
the framework of dark energy, quintessence and other approaches. 

\section{Remarks on fundamental theories}
\label{fundamentaltheories}

Theories dealing with foundation of physics are called fundamental. In this section we discuss some
of those theories and their comparisons. One of the known examples is the comparison 
of nonrelativistic theory (NT) with relativistic one (RT). One of the reasons why RT can be treated as
more fundamental is that it contains a finite parameter $c$ and NT can be treated as a special degenerate
case of RT in the formal limit $c\to\infty$. Therefore, by choosing a large value of $c$,  RT can reproduce any result of NT with a high accuracy. On the contrary, when the limit is already taken one cannot return back from NT to RT and NT cannot reproduce all results of RT. It can reproduce only results obtained when $v\ll c$. Other known examples
are that classical theory is a special degenerate case of quantum one in the formal limit $\hbar\to 0$ and
RT is a special degenerate case of dS and AdS invariant theories in the formal
limit $R\to\infty$ where $R$ is the parameter of contraction from the dS or AdS algebras to the Poincare algebra (see below). A question arises whether it is possible to give a general definition when theory A is more
fundamental than theory B. In view of the above examples, we propose the following

{\bf Definition:} {\it Let theory A contain a finite parameter and theory B be obtained from theory A in the formal limit when the parameter goes to zero or infinity. Suppose that with any desired accuracy theory A can reproduce any result of theory B by choosing a value of the parameter. On the contrary, when the limit is already
taken then one cannot return back to theory A and theory B cannot reproduce all results of theory A. Then theory A is more fundamental than theory B and theory B is a special degenerate case of theory A}. 

A problem arises how to justify this
{\bf Definition} not only from physical but also from mathematical considerations.

In relativistic quantum theory the usual approach to symmetry on quantum level follows. 
Since the Poincare group is the group of motions of Minkowski space, quantum states should be described by representations of this group. 
This implies that the representation generators commute according to the commutation relations of the Poincare group Lie algebra:
\begin{eqnarray}
&&[P^{\mu},P^{\nu}]=0,\quad [P^{\mu},M^{\nu\rho}]=-i(\eta^{\mu\rho}P^{\nu}-
\eta^{\mu\nu}P^{\rho}),\nonumber\\
&&[M^{\mu\nu},M^{\rho\sigma}]=-i (\eta^{\mu\rho}M^{\nu\sigma}+\eta^{\nu\sigma}M^{\mu\rho}-
\eta^{\mu\sigma}M^{\nu\rho}-\eta^{\nu\rho}M^{\mu\sigma})
\label{PCR}
\end{eqnarray}
where $\mu,\nu=0,1,2,3$, $P^{\mu}$ are the operators of the four-momentum and  $M^{\mu\nu}$ are the operators of Lorentz angular momenta. This approach is in the spirit of Klein's Erlangen program in mathematics.

However, background space is only a classical notion which does not exist in quantum theory. For example,
although in QED, QCD and electroweak theory the Lagrangian density 
depends on the four-vector $x$ which is associated with a point in Minkowski space, this is only the
integration parameter which is used in the intermediate stage. The goal of the theory is to construct the
$S$-matrix and when the theory is already constructed
one can forget about Minkowski space because no physical quantity depends on $x$. This is in the
spirit of the Heisenberg $S$-matrix program according to which in relativistic quantum theory it is possible to describe only transitions of states from the infinite past when $t\to -\infty$ to the distant future 
when $t\to +\infty$. For those reasons, as argued in Ref. \cite{PRD}, the approach should be the opposite. Each system is described by a set of linearly  independent operators.
By definition, the rules how they commute with each other define the symmetry algebra. 
In particular, {\it by definition}, Poincare symmetry on quantum level means that the operators commute
according to Eq. (\ref{PCR}). This definition does not involve Minkowski space at all.

Such a definition of symmetry on quantum level has been proposed in Ref. \cite{BKT} and in
subsequent publications of those authors. I am very grateful to Leonid Avksent'evich Kondratyuk for explaining me this definition during our collaboration. I believe that this replacement of the standard paradigm is fundamental for understanding quantum theory, and I did not succeed in finding a similar idea in the literature. 

Our goal is to compare four theories: classical (i.e. non-quantum) theory, nonrelativistic quantum theory, relativistic quantum theory and dS or AdS quantum theory. All those theories are described by representations of the symmetry 
algebra containing ten linearly independent operators $A_{\alpha}\,\, (\alpha=1,2,...10)$: four energy-momentum operators, three angular momentum operators and three Galilei or Lorentz boost operators. For definiteness we assume that the 
operators $A_{\alpha}$ where $\alpha=1,2,3,4$ refer to energy-momentum operators, the operators $A_{\alpha}$ where $\alpha=5,6,7$ refer to angular momentum operators and the operators $A_{\alpha}$ where $\alpha=8,9,10$ refer to Galilei or 
Lorentz boost operators. Let $[A_{\alpha},A_{\beta}]=ic_{\alpha\beta\gamma}A_{\gamma}$ where summation over repeated indices is assumed. In the theory of Lie algebras the quantities $c_{\alpha\beta\gamma}$ are called the structure constants. 

Let $S_0$ be a set 
of $(\alpha,\beta)$ pairs such that $c_{\alpha\beta\gamma}=0$ for all values of $\gamma$ and $S_1$ be a set of $(\alpha,\beta)$ pairs such that $c_{\alpha\beta\gamma}\neq 0$ at least for some values of $\gamma$. Since
$c_{\alpha\beta\gamma}=-c_{\beta\alpha\gamma}$ it suffices to consider only such $(\alpha,\beta)$ pairs
where $\alpha<\beta$. If $(\alpha,\beta)\in S_0$ then the operators $A_{\alpha}$ and $A_{\beta}$ commute
while if $(\alpha,\beta)\in S_1$ then they do not commute.

Let $(S_0^A,S_1^A)$ be the sets $(S_0,S_1)$ for theory A and $(S_0^B,S_1^B)$ be the sets $(S_0,S_1)$ for theory B. As noted above, we will consider only theories where $\alpha,\beta=1,2,...10$. Then one can prove the following 

{\bf Statement:} {\it Let theory A contain a finite parameter and theory B be obtained from theory A in the formal limit when the parameter goes to zero or infinity. If the sets $S_0^A$ and $S_0^B$ are different and $S_0^A 
\subset S_0^B$ (what equivalent to $S_1^B \subset S_1^A$)
then theory A is more fundamental than theory B and theory B is a special degenerate case of theory A.}

Proof: Let ${\tilde S}$ be the set of $(\alpha,\beta)$ pairs such that $(\alpha,\beta)\in S_1^A$ and
$(\alpha,\beta)\in S_0^B$. Then, in theory B, $c_{\alpha\beta\gamma}=0$ for any $\gamma$. One can choose
the parameter such that in theory A all the quantities $c_{\alpha\beta\gamma}$ are arbitrarily small.
Therefore, by choosing a value of the parameter, theory A can reproduce any result of theory B with any
desired accuracy. When the limit is already taken then, in theory B, $[A_{\alpha},A_{\beta}]=0$ for all
$(\alpha,\beta)\in {\tilde S}$. This means that the operators $A_{\alpha}$ and $A_{\beta}$ become
fully independent and therefore there is no way to return to the situation when they do not commute.
Therefore for theories A and B the conditions of {\bf Definition} are satisfied.

It is sometimes stated that the expressions in Eq. (\ref{PCR}) are not general enough because they are
written in the system of units $c=\hbar =1$. Let us consider this problem in more details.
The operators $M^{\mu\nu}$ in Eq. (\ref{PCR}) are dimensionless. In particular, standard angular momentum
operators $(J_x,J_y,J_z)=(M^{12},M^{31},M^{23})$ are dimensionless and satisfy the commutation relations
\begin{equation}
[J_x,J_y]=iJ_z,\quad [J_z,J_x]=iJ_y,\quad [J_y,J_z]=iJ_x
\label{J}
\end{equation}
If one requires that the operators $M^{\mu\nu}$
should have the dimension $kg\cdot m^2/s$ then they should be replaced by $M^{\mu\nu}/\hbar$, respectively.
In that case the new commutation relations will have the same form as in Eqs. (\ref{PCR}) and (\ref{J}) but
the right-hand-sides will contain the additional factor $\hbar$.

The result for the components of angular momentum depends on the system of units. As shown in quantum
theory, in units $\hbar=1$ the result is given by a half-integer
$0, \pm 1/2, \pm 1,...$. We can reverse the order of units
and say that in units where the angular momentum is a half-integer $l$, its
value in $kg\cdot m^2/s$ is $1.0545718\cdot 10^{-34}\cdot
l\cdot kg\cdot  m^2/s$. Which of those two values has more
physical significance? In units where the angular momentum
components are half-integers, the commutation relations (\ref{J})
do not depend on any parameters. Then the meaning of
$l$ is clear: it shows how large the angular momentum is in
comparison with the minimum nonzero value 1/2. At the same time,
the measurement of the angular momentum in units $kg\cdot
m^2/s$ reflects only a historic fact that at macroscopic
conditions on the Earth between the 18th and 21st
centuries people measured the angular momentum in such units.

We conclude that for quantum theory itself the quantity $\hbar$ is not needed. However, it
is needed for the transition from quantum theory to classical one: we introduce $\hbar$, then 
the operators $M^{\mu\nu}$ have the dimension $kg\cdot m^2/s$, and since the right-hand-sides
of Eqs. (\ref{PCR}) and (\ref{J}) in this case contain an additional factor $\hbar$, all the 
commutation relations disappear in the formal limit $\hbar\to 0$. Therefore in classical theory
the set $S_1$ is empty and all the $(\alpha,\beta)$ pairs belong to $S_0$. Since in quantum theory
there exist $(\alpha,\beta)$ pairs such that the operators $A_{\alpha}$ and $A_{\beta}$ do not commute
then in quantum theory the set $S_1$ is not empty and, as follows from {\bf Statement},
classical theory is the special degenerate case of quantum one in the formal limit $\hbar\to 0$.
Since in classical theory all operators commute with each other then in this theory operators are not needed
and one can work only with physical quantities. A question why $\hbar$ is as is does not arise
since the answer is: because people want to measure angular momenta in $kg\cdot m^2/s$.

Consider now the relation between RT and NT. If we introduce the Lorentz boost operators 
$L^j=M^{0j}\,\, (j=1,2,3)$ then Eqs. (\ref{PCR}) can be written as
\begin{eqnarray}
&&[P^0,P^j]=0,\quad [P^j,P^k]=0, \quad [J^j,P^0]=0,\quad [J^j,P^k]=
i\epsilon_{jkl}P^l,\nonumber\\
&&[J^j,J^k]=i\epsilon_{jkl}J^l,\quad [J^j,L^k]=i\epsilon_{jkl}L^l,\quad [L^j,P^0]=iP^j
\label{RT1}
\end{eqnarray}
\begin{equation}
[L^j,P^k]=i\delta_{jk}P^0,\quad [L^j,L^k]=-i\epsilon_{jkl}J^l
\label{RT2}
\end{equation}
where $j,k,l=1,2,3$, $\epsilon_{jkl}$ is the fully asymmetric tensor such that $\epsilon_{123}=1$, 
$\delta_{jk}$ is the Kronecker symbol and a summation over repeated indices is assumed.
If we now define the energy and Galilei boost operators as $E=P^0c$ and $G^j=L^j/c\,\, (j=1,2,3)$,
respectively then the new expressions in Eqs. (\ref{RT1}) will have the same form while instead of
Eq. (\ref{RT2}) we will have
\begin{equation}
[G^j,P^k]=i\delta_{jk}E/c^2,\quad [G^j,G^k]=-i\epsilon_{jkl}J^l/c^2
\label{NT2}
\end{equation}

Note that for relativistic theory itself the quantity $c$ is not needed. In this theory the primary quantities
describing particles are their momenta ${\bf p}$ and energies $E$ while the velocity ${\bf v}$ of a
particle is {\it defined} as ${\bf v}={\bf p}/E$. This definition does not involve meters and seconds, and
the velocities ${\bf v}$ are dimensionless quantities such that 
$|{\bf v}|\leq 1$ if tachyons are not
taken into account. One needs $c$ only for transition from RT to NT: when we introduce $c$ then the 
velocity of a particle becomes ${\bf p}c^2/E$, and its dimension becomes $m/s$. In this case, 
instead of the operators $P^0$ and $L^j$ we work with the operators $E$
and $G^j$, respectively. If $M$ is the Casimir operator for the Poincare algebra defined such that
$M^2c^4=E^2-{\bf P}^2c^2$ then in the formal limit $c\to\infty$ the first expression in Eq. (\ref{NT2})
becomes $[G^j,P^k]=i\delta_{jk}M$ while the commutators in the second expression become zero.
Therefore in NT the $(\alpha,\beta)$ pairs with $\alpha,\beta=8,9,10$ belong to $S_0$ while in RT
they belong to $S_1$. Therefore, as follows from {\bf Statement}, NT is a special degenerate case of RT in 
the formal limit $c\to\infty$. The question why $c=3\cdot 10^8 m/s$ and not, say
$c=7\cdot 10^9 m/s$ does not arise since the answer is: because people want to
measure $c$ in $m/s$.

From the mathematical point of view, $c$ is the parameter of contraction from the Poincare algebra to the Galilei one. This
parameter must be finite: the formal case $c=\infty$ corresponds to the situation when the Poincare algebra does not exist because it becomes the Galilei algebra.

In his famous paper "Missed Opportunities" \cite{Dyson} Dyson notes that RT is more fundamental than NT, and dS and AdS theories
are more fundamental than RT not only from physical but also from pure mathematical considerations. Poincare group is
more symmetric than Galilei one and the transition from the former to the latter at $c\to\infty$ is called contraction. Analogously dS and AdS groups are more symmetric than Poincare one and the transition from the former to the latter at $R\to\infty$ (described below) also is called contraction. At the same time, since dS and AdS groups are semisimple they have a maximum possible symmetry and cannot be obtained from more symmetric groups by contraction.
However, since we treat symmetry not from the point of view of a group of motion for the corresponding background space but from the point of view of commutation relations in the symmetry algebra, we will discuss 
the relations between the dS and AdS algebra on one hand and the Poincare algebra on the other.

By analogy with the definition of Poincare symmetry on quantum level, the definition of dS symmetry on quantum level should not
involve the fact that the dS group is the group of motions of dS space.
Instead, {\it the definition} is that the operators $M^{ab}$ ($a,b=0,1,2,3,4$, $M^{ab}=-M^{ba}$)
describing the system under consideration satisfy the
commutation relations {\it of the dS Lie algebra} so(1,4), {\it i.e.}
\begin{equation}
[M^{ab},M^{cd}]=-i (\eta^{ac}M^{bd}+\eta^{bd}M^{ac}-
\eta^{ad}M^{bc}-\eta^{bc}M^{ad})
\label{CR}
\end{equation}
where $\eta^{ab}$ is the diagonal metric tensor such that
$\eta^{00}=-\eta^{11}=-\eta^{22}=-\eta^{33}=-\eta^{44}=1$.
The {\it definition} of AdS symmetry on quantum level is given by the same equations
but $\eta^{44}=1$.

With such a definition of symmetry on quantum level, dS and AdS
symmetries are more natural than Poincare symmetry. In the
dS and AdS cases all the ten representation operators of the symmetry
algebra are angular momenta while in the Poincare case only six
of them are angular momenta and the remaining four operators
represent standard energy and momentum. If we {\it define} the
operators $P^{\mu}$ as $P^{\mu}=M^{4\mu}/R$ where $R$ is a parameter with the dimension
$length$ then in the formal
limit when $R\to\infty$, $M^{4\mu}\to\infty$ but the quantities
$P^{\mu}$ are finite, Eqs. (\ref{CR}) become Eqs. (\ref{PCR}). This procedure is called contraction and 
in the given case it is the same for the dS or AdS symmetry. As follows from Eqs. (\ref{PCR}) and
(\ref{CR}), if $\alpha,\beta=1,2,3,4$ then the $(\alpha,\beta)$ pairs belong to $S_0$ in
RT and to $S_1$ in dS and AdS theories. Therefore, as follows from {\bf Statement}, 
RT is indeed
a special degenerate case of dS and AdS theories in the formal limit $R\to\infty$. 
By analogy with the abovementioned fact that $c$ must be finite, $R$ must be finite too: the formal case $R=\infty$ corresponds to the 
situation when the dS and AdS algebras do not exist because they become the Poincare algebra.

One of the consequences is that the CC problem described in Sec. \ref{intro} does not exist
because its formulation is based on the incorrect assumption that RT is more fundamental than
dS and AdS theories. Note that the operators in Eq. (\ref{CR}) do not depend on $R$ at all. This
quantity is needed only for transition from dS quantum theory to
Poincare quantum theory. In full analogy with the above discussion of quantities $\hbar$ and $c$, 
a question why $R$ is
as is does not arise and the answer is: because people want to measure distances
in meters. 

On classical level,  dS space is usually treated as the
four-dimensional hypersphere in the five-dimensional space such that
\begin{equation}
x_1^2+x_2^2+x_3^2+x_4^2-x_0^2=R^{'2}
\label{dSspace}
\end{equation}
where $R'$ is the radius of dS space and at this stage it is not clear whether or 
not $R'$ coincide with $R$. 
Transformations from the dS group are usual and hyperbolic rotations of this
space. They can be parametrized by usual and hyperbolic angles and do not
depend on $R'$. In particular, if instead of $x_a$ we introduce the quantities
$\xi_a=x_a/R'$ then the dS space can be represented as a set of points 
\begin{equation}
\xi_1^2+\xi_2^2+\xi_3^2+\xi_4^2-\xi_0^2=1
\label{dSxispace}
\end{equation}
Therefore in classical dS theory itself the quantity $R'$ is not needed at all. It is needed
only for transition from dS space to Minkowski one: we choose $R'$ in meters,
then  the curvature of this space is 
$\Lambda=3/R^{'2}$ and a vicinity of the point $x_4=R'$ or
$x_4=-R'$ becomes Minkowski space in the formal limit $R'\to\infty$.
Analogous remarks are  valid for the transition from AdS theory to Poincare one,
and in this case $\Lambda=-3/R^{'2}$.

We have proved that all the three discussed comparisons satisfy the conditions formulated in {\bf Definition} above.
Namely, the more fundamental theory contains a finite
parameter and the less fundamental theory can be treated as a special degenerate case of the former in the
formal limit when the parameter goes to zero or infinity. The more fundamental theory can reproduce all results of
the less fundamental one by choosing some value of the parameter. On the contrary, when the limit is already taken
one cannot return back from the less fundamental theory to the more fundamental one. 

In Ref. \cite{JPA} we considered properties of dS quantum theory and argued that
dS symmetry is more natural than Poincare one. However, the above discussion proves that dS
and AdS symmetries are not only more natural than Poincare symmetry but more fundamental.
In particular, $R$ is fundamental to the same extent as $\hbar$ and $c$ and, as noted above, $R$ {\bf must be finite}.

\section{A system of two bodies in quantum dS theory}
\label{2bodies}

Let us stress that the above proof
that dS symmetry is more fundamental than Poincare one has been performed on pure quantum
level. In particular, the proof does not involve the notion of background space 
and
the notion of $\Lambda$. Therefore a problem arises whether this result can be used for
explaining that experimental data can be described in the framework of GR with $\Lambda>0$.

Our goal is to show that in quantum mechanics based on the dS algebra, classical equations of 
motions for a system of two free macroscopic bodies follow from quantum mechanics in
semiclassical approximation and those equations are the same as in GR with dS background space. 
We will assume that the distance between the
bodies is much greater than the sizes of the bodies and the bodies do not
have anomalously large internal angular momenta. Then from the formal
point of view the motion of two bodies as a whole can be described by the same formulas as the motion of two elementary particles with zero spin. 
In quantum dS theory elementary particles are described by irreducible representations (IRs) of the dS algebra and, as shown in Ref. \cite{JPA},  one can explicitly construct such IRs. 

It is known that in Poincare theory any massive IR can be implemented in the 
Hilbert space of functions $\chi({\bf v})$ on
the Lorenz 4-velocity hyperboloid with the points $v=(v_0,{\bf v}),\,\, v_0=(1+{\bf v}^2)^{1/2}$ such that 
$\int\nolimits |\chi({\bf v})|^2d\rho({\bf v}) <\infty$ and $d\rho({\bf v})=d^3{\bf v}/v_0$ is the Lorenz
invariant volume element. For positive energy IRs the value of energy is $E=mv_0$
where $m$ is the particle mass {\it defined as the positive square root} $(E^2-{\bf P}^2)^{1/2}$.
Therefore for massive IRs, $m>0$ by definition.

However, as shown by Mensky \cite{Mensky}, in contrast to Poincare theory, IRs in dS theory can be implemented only on two Lorenz hyperboloids,
i.e. Hilbert spaces for such IRs consist of sets of two functions $(\chi_1({\bf v}),\chi_2({\bf v}))$ such that 
$$\int\nolimits (|\chi_1({\bf v})|^2+|\chi_2({\bf v})|^2)d\rho({\bf v}) <\infty$$
In Poincare limit one dS IR splits into two IRs of the Poincare algebra with positive and negative energies and, as argued in Ref. \cite{JPA}, this implies that one IR of the dS algebra describes a particle and its
antiparticle simultaneously. Since in the present paper we do not deal with antiparticles and neglect spin
effects, we give
only expressions for the action of the operators on the upper hyperboloid in the case of zero spin \cite{JPA}:
\begin{eqnarray}
&&{\bf J}=l({\bf v}),\quad {\bf L}=-i v_0 \frac{\partial}{\partial {\bf v}},\quad {\bf B}=m_{dS} {\bf v}+i [\frac{\partial}{\partial {\bf v}}+
{\bf v}({\bf v}\frac{\partial}{\partial {\bf v}})+\frac{3}{2}{\bf v}]\nonumber\\
&& {\cal E}=m_{dS} v_0+i v_0({\bf v}
\frac{\partial}{\partial {\bf v}}+\frac{3}{2})
\label{IR1}
\end{eqnarray}
where ${\bf B}=\{M^{41},M^{42},M^{43}\}$, ${\bf l}({\bf v})=-i{\bf v}\times \partial/\partial {\bf
v}$, ${\cal E}=M^{40}$ and $m_{dS}$ is a positive quantity. 

This implementation of the IR is convenient for the transition to Poincare limit. Indeed, the operators
of the Lorenz algebra in Eq. (\ref{IR1}) are the same as in the IR of the Poincare algebra. Suppose
that the limit of $m_{dS}/R$ when $R\to\infty$ is finite and denote this limit as $m$. Then in the
limit $R\to\infty$ we get standard expressions for the operators of the IR of the Poincare algebra
where $m$ is standard mass, $E={\cal E}/R=mv_0$ and ${\bf P}={\bf B}/R=m{\bf v}$. For this
reason $m_{dS}$ has the meaning of the dS mass. 
Since Poincare symmetry is a special case of dS one, $m_{dS}$ is more fundamental than $m$.
Since Poincare symmetry works with a high accuracy, the value of $R$ is supposed to be very large (but, as noted above, it cannot be infinite).

Consider the non-relativistic approximation when $|{\bf v}|\ll
1$. If we wish to work with units where the dimension of
velocity is $meter/sec$, we should replace ${\bf v}$ by ${\bf
v}/c$. If ${\bf p}=m{\bf v}$ then it is clear from the
expression for ${\bf B}$ in Eq. (\ref{IR1}) that ${\bf p}$ becomes the real momentum ${\bf P}$
only in the limit $R\to\infty$. At this
stage we do not have any coordinate space yet. However, by analogy with
standard quantum mechanics, we can {\it define} the position
operator ${\bf r}$ as $i\partial/\partial {\bf p}$. 

In classical approximation we can treat
${\bf p}$ and ${\bf r}$ as usual vectors. Then as follows from Eq. (\ref{IR1})
\begin{equation}
{\bf P}= {\bf p}+mc{\bf r}/R, \quad H = {\bf p}^2/2m +c{\bf p}{\bf r}/R,\quad {\bf L}=-m{\bf r}
\label{PH}
\end{equation}
where $H=E-mc^2$ is the classical nonrelativistic Hamiltonian. As follows from these expressions, 
\begin{equation}
H({\bf P},{\bf r})=\frac{{\bf P}^2}{2m}-\frac{mc^2{\bf r}^2}{2R^2}
\label{HP}
\end{equation}

The last term in Eq. (\ref{HP}) is the dS correction to
the non-relativistic Hamiltonian. It is interesting to note
that the non-relativistic Hamiltonian depends on $c$ although
it is usually believed that $c$ can be present only in
relativistic theory. This illustrates the fact mentioned in Sec. \ref{fundamentaltheories}
that the transition to nonrelativistic theory
understood as $|{\bf v}|\ll 1$ is more physical than that
understood as $c\to\infty$. The presence of $c$ in Eq.
(\ref{HP}) is a consequence of the fact that this expression is
written in standard units. In nonrelativistic theory $c$ is
usually treated as a very large quantity. Nevertheless, the
last term in Eq. (\ref{HP}) is not large since we assume
that $R$ is very large.

As follows from Eq. (\ref{HP}) and the Hamilton equations, in dS theory a free particle
moves with the acceleration given by
\begin{equation}
{\bf a}={\bf r}c^2/R^2
\label{accel}
\end{equation}
where ${\bf a}$ and ${\bf r}$ are the acceleration and the radius vector of the particle, respectively.
Since $R$ is very large, the acceleration is not negligible only at cosmological distances
when $|{\bf r}|$ is of the order of $R$. 

Following our results in Ref. \cite{JPA}, we now consider whether the result (\ref{accel}) is compatible with GR. 
As noted in Sec. \ref{fundamentaltheories}, the dS space is a four-dimensional manifold in the five-dimensional space defined by Eq. (\ref{dSspace}).
In the formal limit $R'\to\infty$ the action of the dS group in
a vicinity of the point $(0,0,0,0,x_4=R')$ becomes the action
of the Poincare group on Minkowski space. With this parameterization
 the metric tensor on dS space is
\begin{equation}
g_{\mu\nu}=\eta_{\mu\nu}-x_{\mu}x_{\nu}/(R^{'2}+x_{\rho}x^{\rho})
\label{metric}
\end{equation}
where $\mu,\nu,\rho = 0,1,2,3$, $\eta_{\mu\nu}$ is the Minkowski metric tensor,  
and a summation
over repeated indices is assumed. It is easy to calculate the
Christoffel symbols in the approximation where all the
components of the vector $x$ are much less than $R'$:
$\Gamma_{\mu,\nu\rho}=-x_{\mu}\eta_{\nu\rho}/R^{'2}$. Then a
direct calculation shows that in the nonrelativistic
approximation the equation of motion for a single particle is
 the same as in Eq. (\ref{accel}) if $R'=R$.

Another way to show that Eq. (\ref{accel}) is compatible with GR follows. The known result of GR is that if the metric
is stationary and differs slightly from the Minkowskian one
then in the nonrelativistic approximation the curved space-time
can be effectively described by a gravitational potential
$\varphi({\bf r})=(g_{00}({\bf r})-1)/2c^2$. We now express
$x_0$ in Eq. (\ref{dSspace}) in terms of a new variable $t$ as
$x_0=t+t^3/6R^{'2}-t{\bf x}^2/2R^{'2}$. Then the expression for the
interval becomes
\begin{equation}
ds^2=dt^2(1-{\bf r}^2/R^{'2})-d{\bf r}^2-
({\bf r}d{\bf r}/R')^2
\label{II67}
\end{equation}
Therefore, the metric becomes stationary and $\varphi({\bf r})=-{\bf r}^2/2R^{'2}$ in agreement with Eq. (\ref{accel}) if $R'=R$. The fact that in classical approximation the parameter $R$ defining contraction from quantum dS
symmetry to quantum Poincare symmetry becomes equal the radius of dS space in GR does not mean
that $R$ can be always identified with this radius because on quantum level the notion of background space does not have the physical meaning.

Consider now a system of two free classical bodies in GR. Let $({\bf r}_i,{\bf a}_i)$
$(i=1,2)$ be their radius vectors and accelerations, respectively. Then Eq. (\ref{accel}) is
valid for each particle if $({\bf r},{\bf a})$ is replaced by $({\bf r}_i,{\bf a}_i)$, respectively.
Now if we define the relative radius vector ${\bf r}={\bf r}_1-{\bf r}_2$ and the 
relative acceleration ${\bf a}={\bf a}_1-{\bf a}_2$ then they will satisfy the same Eq. (\ref{accel})
which shows that the dS antigravity is repulsive. 

Let us now consider a system of two free bodies in the framework of the representation of the dS algebra. The particles are described by the
variables ${\bf P}_j$ and ${\bf r}_j$ ($j=1,2$). Define standard nonrelativistic variables
\begin{eqnarray}
&&{\bf P}_{12}={\bf P}_1+{\bf P}_2,
\quad {\bf q}=(m_2{\bf P}_1-m_1{\bf P}_2)/(m_1+m_2)\nonumber\\
&&{\bf R}_{12}=(m_1{\bf r}_1+m_2{\bf r}_2)/(m_1+m_2),\quad
{\bf r}={\bf r}_1-{\bf r}_2
\label{2body}
\end{eqnarray}
Then, as follows from Eq. (\ref{PH}), in the
nonrelativistic approximation the two-particle quantities ${\bf P}$, ${\bf E}$ and ${\bf L}$ are given by
\begin{equation}
{\bf P}= {\bf P}_{12},\quad E = M+\frac{{\bf P}_{12}^2}{2M} -\frac{Mc^2{\bf R}_{12}^2}{2R^2},
\quad {\bf L}=-M{\bf R}_{12}\label{2PE}
\end{equation}
where
\begin{equation}
M = M({\bf q},{\bf r})=
m_1+m_2 +H_{nr}({\bf r},{\bf q}),\quad 
H_{nr}({\bf r},{\bf q})=\frac{{\bf q}^2}{2m_{12}}-\frac{m_{12}c^2{\bf r}^2}{2R^2}
\label{2M}
\end{equation}
and $m_{12}$ is the reduced two-particle mass. Here the operator $M$ acts in the space of functions
$\chi({\bf q})$ such that $\int |\chi({\bf q})|^2d^3{\bf q}<\infty$ and ${\bf r}$ acts in this space as
${\bf r}=i\partial/\partial {\bf q}$. 

It now follows from Eq. (\ref{IR1}) that $M$ has the meaning of the two-body mass and
therefore $M({\bf q},{\bf r})$ is the internal two-body Hamiltonian. Then, by analogy with the derivation of Eq. (\ref{accel}),
it can be shown from the Hamilton equations that in semiclassical approximation the relative
acceleration is given by the same expression (\ref{accel}) but now
${\bf a}$ is the relative acceleration and ${\bf r}$ is the
relative radius vector.

\section{To what extent are the quantities $(c,\hbar,s)$  fundamental?}
\label{units}
 
In the literature the notion of the $c\hbar G$ cube of physical theories is sometimes used. The meaning is
that any relativistic theory should contain $c$, any quantum theory should contain $\hbar$ and any gravitation 
theory should contain $G$.
The more fundamental a theory is the greater number of those parameters it contains. In particular, relativistic quantum theory of
gravity is treated as the most fundamental because it contains all the three parameters $c$, $\hbar$ and $G$ while nonrelativistic
classical theory without gravitation is the least fundamental because it contains none of those parameters.

However, as noted in Sec. \ref{intro}, since the nature of gravity is not clear yet,
the quantity $G$ is not fundamental. Also, as follows from the above discussion, the set of parameters $(c,\hbar,R)$
is more adequate than the set $(c,\hbar,G)$ and, in contrast to usual statements,
the situation is the opposite: relativistic theory should not contain $c$, quantum theory should not contain $\hbar$ and dS or AdS theories should not contain $R$.
Those three parameters are needed only for transitions from more general
theories to less general ones. The most general dS and AdS quantum theories
do not contain dimensionful quantities at all while the least general nonrelativistic
classical theory contains three dimensionful quantities $(kg,m,s)$.

Indeed, as noted above, the angular momenta are dimensionless but for historical reasons people want to measure them in $kg\cdot m^2/s$ and that's why 
the quantity $\hbar$ arises. Analogously, in particle theory, velocities are dimensionless but since people want to measure them in $m/s$ the quantity $c$ comes into play.
However, when a system under consideration is strongly quantum and Poincare symmetry does not work, neither
the quantities $(kg,m,s)$ nor the quantities $(c,\hbar,R)$ have a physical meaning and those quantities are not present in the theory at all.

Nevertheless, physicists usually believe that the quantities $(c,\hbar)$ are fundamental and
do not change over time. This belief has been implemented in the modern system of units where the basic quantities are not $(kg,m,s)$ but $(c,\hbar,s)$ and it is postulated
that the quantities $(c,\hbar)$ do not change over time. By definition, it is postulated that from now on $c=299792458m/s$ and $\hbar=1.054571800\cdot 10^{-34}kg\cdot m^2/s$. As a consequence, now the quantities $(kg,m)$ are
not basic ones because they can be expressed in terms of $(c,\hbar,s)$ while the second remains the basic quantity.

The motivation for the modern system of units is based on several facts of quantum theory based on Poincare invariance. First of all, since it is
postulated that the photon is massless, its speed $c$ is always the same for any photons with any energies. Another postulate is that for any photon its energy is
always proportional to its frequency and the coefficient of proportionality always equals $\hbar$. Let us note that this terminology might be misleading for the following reasons.
Since the photon is the massless  elementary particle, it is characterized only by energy, momentum, spin and helicity and {\it is not characterized} by frequency and wave length. The latter are only classical notions characterizing
a classical electromagnetic wave containing many photons. Quantum theory predicts the energy distribution of photons in blackbody radiation  but experimentally we cannot follow individual photons and can measure only the
frequency distribution in the radiation. Then the theory agrees with experiment if formally the photon with the
energy $E$ is attributed the frequency $\omega=E/\hbar$. 

A typical theoretical justification is that the photon wave function contains $exp(-iEt/\hbar)$. Note that the problem of time is one of the most fundamental problems of standard quantum theory. The usual point of view is that there is no time operator and time is simply a classical parameter such that the wave function contains the factor $exp(-iEt/\hbar)$
(see e.g. the discussion in Ref. \cite{lev5}). This agrees with the facts that in classical approximation the Schr\"{o}dinger
equation becomes the Hamilton-Jacobi equations and that with such a dependence of the wave function on time
one can describe trajectories of photons in classical approximation (see e.g. the discussion in Ref. \cite{lev6}).
At the same time, there is no experimental proof that this dependence takes place on quantum level and, as noted in Sec. \ref{fundamentaltheories}, fundamental quantum theories proceed from the Heisenberg S-matrix program  that in quantum theory one can describe only transitions of states from the infinite past when $t\to -\infty$ to the distant future when $t\to +\infty$.  

Consider now the description of the photon in AdS and dS quantum theories but first let us make the following
remarks. Dyson's paper \cite{Dyson} explaining why de Sitter symmetries are more fundamental than
Poincare symmetry appeared in 1972. One might think that this paper should be a good stimulus for
physicists to generalize fundamental quantum theories from Poincare
invariant theories to de Sitter invariant ones. However, no big steps in this direction have been made.
One of the arguments is that since $R$ is much greater than sizes of elementary particles then de Sitter
corrections will be negligible. However, as explained below, in de Sitter and Poincare invariant theories the structures of IRs 
describing elementary particles are considerably different. The analogy is that relativistic theory cannot be treated simply as nonrelativistic one
with the cutoff $c$ for velocities: as a consequence of the fact that $c$ is finite the theories considerably differ in several aspects.

Consider first IRs of the AdS algebra. For the first time the construction of such IRs has been 
given by Evans \cite{Evans} (see also Ref. \cite{tmf}). As noted above, the AdS analog of the 
energy operator is $M^{04}$. A common feature of the AdS and
Poincare cases is that there are IRs containing either only positive or only negative energies and the latter can be associated with antiparticles. In the AdS case the
minimum value of the energy in IRs with positive energies can be treated as the mass by analogy with the Poincare case. However, the essential difference between
the AdS and Poincare cases is that the IRs in the former belong to the discrete series of IRs and the photon mass cannot be exactly zero. In the AdS analog of massless Poincare
IR, the AdS mass equals $m_{AdS}=1+s$ where $s$ is the spin. From the point of view of Poincare symmetry, this is an extremely small quantity since the Poincare mass
$m$ equals $m_{AdS}/R$. However, since $m_{AdS}$ is not exactly zero, there is
a nonzero probability that the photon can be even in the rest state, i.e. its speed will be zero. In general, the speed of the photon can be in the range
$[0,1)$. Therefore, in contrast to Poincare case, there is no situation when all photons with all energies have the same speed. As a consequence, the constant $c$ does not have the fundamental meaning as in Poincare theory.

In addition, as a consequence of the fact that AdS analogs of massless IRs contain the rest state, particles described by such IRs necessarily have two values
of helicity, not one as in Poincare case. Note that in Poincare theory the photon is not described by an IR of the pure Poincare algebra because it is massless and has two
helicities: it is described by an IR of the Poincare algebra with spatial reflection added. For example, if in Poincare theory neutrino is treated as massless then in
AdS theory it will have two helicities. However, if its AdS energy is much greater than its AdS mass then the probability to have the second helicity is very small (but not zero).

Consider now IRs of the dS algebra. They have been constructed in Ref. \cite{JPA} by using the results on IRs of the dS group in the excellent book by Mensky \cite{Mensky}. Here the situation drastically differs from the Poincare case because there are no IRs with only positive and negative energies: one
IR necessarily contains both positive and negative energies. As argued in Ref. \cite{JPA}, this implies that a particle and its antiparticle belong to the same IR. This means that the
very notion of particles and antiparticles is only approximate and the conservation of electric charge and baryon and lepton quantum numbers also is only approximate because
transitions particle$\leftrightarrow$artiparticle are not strongly prohibited. One IR of the dS algebra splits into two IRs of the Poincare algebra in the formal limit $R\to\infty$. IRs of the dS
algebra are characterized by the dS mass $m_{dS}$ such that $m_{dS}$ cannot be zero and the relation between dS and Poincare masses is again $m_{dS}=Rm$. So even the photon is necessarily
massive. In Poincare theory there is a discussion what is the upper bound for the photon mass and different authors give the values in the range $(10^{-17}ev,10^{-14}ev)$. These seem to be
extremely tiny quantities but even if $m=10^{-17}ev$ and $R$ is of the order of $10^{26}meters$ as usually accepted than $m_{dS}$ is of the order of $10^{16}$, i.e. a very large quantity.
We conclude that in the dS case the quantity $c$ cannot have a fundamental meaning, as well as in the AdS case.

Consider now whether the quantity $\hbar$ can be treated as fundamental in de Sitter invariant theories.  For such theories
it is not even clear how to define energy and time such that the wave function depends on time 
as $exp(-iEt/\hbar)$ even in classical approximation. For example, in the dS case the operator $M^{04}$ 
is on the same footing as the operators
$M^{0j}$ ($j=1,2,3$) and only in Poincare limit it becomes the energy operator. In Sec. \ref{2bodies} this problem is solved in the approximation $1/R^2$ but in the general case the problem remains open. 

While in the modern system of units, $c$ and $\hbar$ are treated as exact quantities, the second is treated only as
an approximate quantity. Since there is no time operator, it is not even legitimate to say whether time should
be discrete or continuous. The second is defined as the duration of 9192631770 periods of the radiation corresponding to the transition between the two hyperfine levels of the ground state of the cesium 133 atom. The
physical quantity describing the transition is the transition energy $\Delta E$, and the frequency of the radiation is defined as $\Delta E/\hbar$. The transition energy cannot be the exact quantity because the width of the
transition energies cannot be zero. In addition, the transition energy depends on 
 gravitational and electromagnetic fields and on other phenomena. In view of all those phenomena the accuracy of one second given in the literature
is in the range $(10^{-18}s,10^{-16}s)$, and the better accuracy cannot be obtained in principle. 
In summary, "continuous time" is a part of classical notion of space-time continuum and makes no sense beyond this notion.

In modern inflationary models  the inflation period of the Universe lasted in the range $(10^{-36}s,10^{-32}s)$ after the Big Bang. In addition to the fact that such times cannot be measured in principle, at this stage of the Universe
there were no nuclei and atoms and so it is unclear whether time can be defined at all. The philosophy of classical physics is that any physical quantity can be measured with any desired accuracy. However the state of the Universe at that time
could not be classical, and in quantum theory the definition of any physical quantity is a description how this
quantity can be measured, at least in principle. In quantum theory it is not acceptable to say that "in fact" some 
quantity exists but cannot be measured. So in our opinion, description of the inflationary period by times
$(10^{-36}s,10^{-32}s)$ has no physical meaning. 

In summary, since in dS and AdS theories all physical quantities are dimensionless, here no system of units
is needed. Dimensionful quantities $(c,\hbar,s)$ are meaningful only at special conditions when Poincare
symmetry works with a high accuracy and measurements can be performed in a system which is
classical (i.e. non-quantum) with a high accuracy.

\section{Conclusion}
\label{conclusion}

In Sec. \ref{fundamentaltheories} we have proved that dS and AdS quantum theories are more fundamental
than Poincare quantum theory. The transition from the former to the latter is described by contraction
$R\to\infty$. The parameter $R$ has nothing to do with the radius of background space and must be finite. 
As shown in Sec. \ref{2bodies}, as a consequence of those results, in semiclassical approximation two free bodies have a relative acceleration defined by the same expression as in GR if the the radius of dS space equals $R$ and $\Lambda=3/R^2$. 
This result has been obtained without using dS space, its
metric, connection etc.: it is simply a consequence of dS quantum mechanics of two free bodies
and {\it the calculation does not involve any geometry}. In our opinion this result is more important than the
result of GR because any classical result should be a consequence
of quantum theory in semiclassical approximation.

Therefore, as follows from basic principles of quantum theory, correct description of nature in GR  
implies that $\Lambda$ {\it must} be nonzero, and the problem why $\Lambda$ is as is does not arise.
This has nothing to do with gravity, existence or nonexistence of dark energy and with the problem whether
or not  empty space-time should be necessarily flat. 

As argued in Sec. \ref{units}, since all physical quantities in dS and AdS quantum theories are
dimensionless, here no system of units is needed. The quantities $(c,\hbar,s)$ (which are basic ones in the modern system of units)
have a physical meaning
only at special conditions when Poincare symmetry works with a high accuracy and measurements are
performed in a system which is classical (i.e. non-quantum) with a high accuracy. In particular, 
statements that the inflationary stage of the Universe lasted in the range $(10^{-36}s,10^{-32}s)$ have
no physical meaning.

{\it Acknowledgement:} I am grateful to Bernard Bakker and Vladimir Karmanov for numerous important discussions.

\end{document}